\documentclass[floatfix,superscriptaddress,twocolumn,prl]{revtex4}

\usepackage{amsmath,amssymb}
\usepackage{graphicx}

\usepackage{bm}% bold math

\newcommand{\degree}{{$^\circ$}}
\newcommand{\monolayer}{Sr$_2$RuO$_4$}
\newcommand{\bilayer}{Sr$_3$Ru$_2$O$_7$}
\newcommand{\inflayer}{SrRuO$_3$}

\begin{document}

%\preprint{APS/}

\title{Metamagnetic Quantum Criticality Revealed by $^{17}$O-NMR in the Itinerant Metamagnet Sr$_3$Ru$_2$O$_7$}

\author{K.~Kitagawa}
\affiliation{Department of Physics, Graduate School of Science, Kyoto
University, Kyoto 606-8502, Japan}

\author{K.~Ishida}
\affiliation{Department of Physics, Graduate School of Science, Kyoto
University, Kyoto 606-8502, Japan}

\author{R.~S.~Perry}
\altaffiliation{School of Physics \& Astronomy, University of St. Andrews, North Haugh, St. Andrews, Fife KY16 955, UK.}
\affiliation{Department of Physics, Graduate School of Science, Kyoto University, Kyoto 606-8502, Japan}
\affiliation{International Innovation Center, Kyoto University, Kyoto 606-8501, Japan}

\author{T.~Tayama}
\affiliation{Institute for Solid State Physics, the University of Tokyo, Kashiwa, Chiba 277-8581, Japan}

\author{T.~Sakakibara}
\affiliation{Institute for Solid State Physics, the University of Tokyo, Kashiwa, Chiba 277-8581, Japan}

\author{Y.~Maeno}
\affiliation{Department of Physics, Graduate School of Science, Kyoto University, Kyoto 606-8502, Japan}
\affiliation{International Innovation Center, Kyoto University, Kyoto 606-8501, Japan}

\date{\today}% It is always \today, today,

\begin{abstract}
We have investigated the spin dynamics in the bilayered perovskite Sr$_3$Ru$_2$O$_7$ as a function of magnetic field and temperature using $^{17}$O-NMR. 
This system sits close to a metamagnetic quantum critical point (MMQCP) for the field perpendicular to the ruthenium oxide planes. We confirm Fermi-liquid behavior at low temperatures except for a narrow field region close to the MMQCP. The nuclear spin-lattice relaxation rate divided by temperature $1/T_1T$ is enhanced on approaching the metamagnetic critical field of $\sim 7.9$~T and at the critical field $1/T_1T$ continues to increase and does not show Fermi- liquid behavior down to 0.3~K.  The temperature dependence of $T_1T$ in this region suggests the critical temperature $\mathit{\Theta}$ to be $\sim 0$~K, which is a strong evidence that the spin dynamics possesses a quantum critical character. Comparison between uniform susceptibility and $1/T_1T$ reveals that antiferromagnetic fluctuations instead of two-dimensional ferromagnetic fluctuations dominate the spin fluctuation spectrum at the critical field, which is unexpected for itinerant metamagnetism. 
\end{abstract}

%Use showkeys class option if keyword
\maketitle

%\section{Introduction}
 Quantum critical points (QCP) occur at a zero-temperature continuous phase transition  where the properties are dominated by quantum as opposed to classical or thermal fluctuations \cite{QPTBook}. Generally, quantum criticality is induced using a non-thermal tuning parameter, most notably pressure, to suppress a second order phase transition to zero temperature. However, this adds considerable technological constraints to experimentalists. Recently, magnetic-field tuned QCPs have become popular because this constraint does not apply. One system recently shown to have such a QCP is the bilayered perovskite ruthenate \bilayer\ \cite{GrigeraQCP}.

In layered perovskite ruthenates ruthenium 4$d$ orbitals hybridize with oxygen 2$p$ orbitals  to create narrow bands and the resulting strong electronic correlations induce various intriguing properties of magnetism and superconductivity. The highest profile member of the series, the single layered \monolayer\ is an unconventional superconductor with most likely spin triplet pairing \cite{214maeno,214review}. The pseudo-cubic \inflayer, on the other hand, is an itinerant ferromagnet \cite{113}.  The bilayered ruthenate \bilayer\ has an intermediate dimensionality to those of \monolayer\ and \inflayer, and as such is situated very close to a ferromagnetic instability \cite{SIkedaPressure}.  At zero applied field, it is a strongly correlated Fermi liquid \cite{SIkeda}.  Under magnetic field, itinerant electron metamagnetism, which is empirically defined as a super-linear rise in the magnetization $M$, is observed \cite{PerryPRL}.  Classically, such metamagnetism originates from the Fermi level sitting close to a peak in the density of states.  On application of a magnetic field $H$, the Fermi surface undergoes a field-induced Stoner transition to a high field polarised state.  Thus the phase diagram consists of a line of first order phase transitions terminating at a critical end point.  Metamagnetic quantum criticality (MMQC) and metamagnetic quantum critical point (MMQCP) are realized when the critical end point is situated at zero temperature \cite{MillisMMQCP,GrigeraQCP}.  This contrasts to many pressure induced QCPs because the MMQCP is non-symmetry breaking. 

The metamagnetism in \bilayer\ is sensitive to the field angle to the ruthenium oxide planes (the $ab$ plane) and the critical end point ($H^{*}$, $T^{*}$) can be tuned by field direction. The end points for various directions are reported \cite{GrigeraPRB}: (5.6~T, 1.2~K) for $H \perp c$ and (7.9~T, 0~K) for $H \parallel c$, therefore \bilayer\ is considered to be close to the MMQCP when the field of 7.9~T is applied along the $c$ axis.  Here, we focus only on the spin dynamics for $H \parallel c$.  An unusual anomaly was also discovered close to the MMQCP for $H \parallel c$ when sample quality was improved significantly.  High-quality samples (residual in-plane resistivity $\rho_0 < 1$~$\mu\Omega\cdot$cm) show at least two jumps in $M(H)$ when field is increased across the metamagnetic field $H_\text{M}$.  In the field range between two jumps (7.85~T$ < H < 8.05$~T), $\rho_0$ is around twice as high as below and above the jumps \cite{PerrySdH} and both $\rho(T)$ and $M(T)$ show unusual features at $\sim 1$~K  \cite{PerrySdH,GrigeraDisorder}.  These results suggest the possible existence of a novel phase in the critical region  \cite{GrigeraDisorder}.  So far, evidence for quantum critical behavior and the novel anomaly in \bilayer\ have only come from bulk measurements.  In this letter, we report the first microscopic evidence of quantum critical fluctuations in this material via a $^{17}$O-NMR study to probe the spin dynamics in magnetic fields up to 14~T and temperatures as low as 0.3~K.

%\section{Experiment}
%sample
The single crystals used in our NMR experiment were grown using a floating-zone technique  \cite{PerryGrowth}.  In order to exchange the natural oxygen for $^{17}$O (nuclear spin $I = 5/2$) annealing was performed in a silica tube furnace for a week at 1050\degree C with 70\% enriched $^{17}$O$_2$ gas. The annealed crystals had a $\rho_0 = 0.8$~$\mu\Omega \cdot$cm; similar in quality to those that displayed both de Haas-van Alphen (dHvA) oscillations  \cite{BorzidHvA} and the putative novel phase \cite{GrigeraDisorder}.  

\begin{figure}[htbp]
  \includegraphics[width=\linewidth,clip]{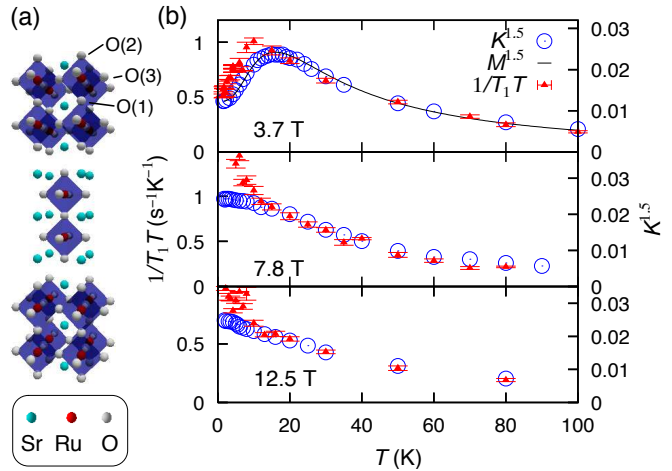} 
%  \vskip -0.05\linewidth
\caption{\label{fig:fmcomparison} (color online)
(a)~The crystal structure of \bilayer. 
(b)~Comparison between $1/T_1T$ of O(2) site and Knight shifts $^{17}K$ of O(3) site below (top: 3.7~T), at (middle: 7.8~T), and above the metamagnetic field (bottom: 12.5~T). The solid line at 3.7~T is magnetization measured by a commercial SQUID magnetometer.
}
\end{figure}

%NMR
There are three inequivalent crystallographic oxygen sites in the bilayered perovskite 
structure as shown in Fig.~\ref{fig:fmcomparison}(a).  We define the O(1), O(2), and O(3) sites as inner-apical, outer- apical, and in-plane sites, respectively. $^{17}$O-NMR spectrum consists of well-separated lines due to the electric quadrupole interaction and difference of the Knight shifts among the sites. The details of the NMR-spectrum analyses will be summarized in a separate paper \cite{KitagawaUnpub}. The O(2) spectrum, which is less affected by the Ru-4$d$ spin than the other two sites, was observed at all temperatures and applied fields, while the $^{17}$O-NMR signals from the O(1) and O(3) sites could not be detected in the vicinity of MMQCP due to large relaxation rates. In this paper we concentrate on analysing the spin dynamics extracted from the NMR at the O(2) site.  Away from the critical field we have confirmed that the microscopic local magnetization deduced from the NMR shift at the three sites is well scaled to the macroscopic bulk $M$. The jump in $M(H)$ is also observed in the NMR shift of the O(2) site at 0.3~K \cite{KitagawaUnpub}. Hence, a relaxation measurement at the O(2) site should reflect the intrinsic nature of the spin dynamics of \bilayer. 

In this study $1/T_1$ was measured by the saturation recovery method and determined with a single component over the whole $H$-$T$ range by fitting a standard theoretical relaxation function \cite{NarathTiNMR}. For a relaxation measurement near the MMQCP at low temperatures where $1/T_1T$ is enhanced, it requires special care to measure a reliable $1/T_1$, due to the eddy current heating induced by the RF pulses.  At low temperatures, the relaxation time of nuclear spins by electron spins is comparable with the thermal equilibrium time after saturation pulses and so in order to check the effect of RF heating we applied off-resonance saturation pulses. In addition, we reduced the heating effect as much as possible by employing low-power, long-width saturation pulses when needed. Thus it took very long time, sometimes 2 days, to obtain one relaxation measurement in the vicinity of the MMQCP.

%\section{Result \& Discussion}
The nuclear spin-lattice relaxation rate $1/T_1$ is related to the wave vector $\bm q$-averaged dynamical susceptibility $\chi''(\bm q,\omega_\text{res})$. The O(2) site is located directly above the Ru site so that $\bm q$ dependence of the hyperfine coupling is likely to be negligible. Thus $1/T_1T$ is expressed as 
 \begin{equation*}
 1/T_1 T \propto \sum_{\bm q} \frac{\chi''(\bm q, \omega_\text{res})}{\omega_\text{res}},
 \end{equation*}
 where $\omega_\text{res}$ is the NMR frequency.  We used a frequency range between 20 MHz and 80 MHz which corresponds to a submilli-Kelvin energy scale.  Comparisons between $1/T_1T$ and uniform susceptibility $\chi(\bm q = \bm 0) = dM/dH$ can give information on the ${\bm q}$-dependence of spin fluctuations.  The NMR Knight shift $K$ which gives microscopic information on $M/H$ can be used instead of $dM/dH$ for such analysis.  For example, when ferromagnetic (FM) fluctuations are dominant, the dynamical susceptibility has a peak around $\bm q \sim 0$ and $1/T_1T$ is related to $K$ and the bulk susceptibility \cite{113NMR}. 

%%%%%%%%%%%%%%%%%%%%% Fig1%%%%%%%%%%%%%%%%%%%%%%%%%%%%%%%%%%%%%%%%% 
%%%%%%%%%%%%%%%%%%%%%%%%%%%%%%%%%%%%%%%%%%%%%%%%%%%%%%%%%%%%%%%%%%%
We start by discussing data in the high temperature region of the phase diagram ($T > 2$~K). The spin dynamics in zero field have been intensively studied by the inelastic neutron scattering (INS) measurements \cite{CapognaINS} and it was revealed that the existence of the two-dimensional (2D) FM fluctuations dominate the properties above 20~K. When the 2D FM fluctuations are significant, $1/T_1T$ can be expressed as,
%  \begin{equation*}
$
 1/T_1T \propto \chi(\bm q=\bm 0)^{3/2}
$ \cite{SCRBook}.
%  \end{equation*}
 In Fig.~\ref{fig:fmcomparison}(b), we show the temperature dependence of $1/T_1T$ measured at the O(2) site and the Knight shift $K^{3/2}$ of the O(3) site scaled with $M^{3/2}$. At 3.7~T, $1/T_1T$ and $K$ have a maximum around 15~K, in agreement with bulk susceptibility measurements \cite{SIkeda}. Above 15~K, $1/T_1T$ shows good scaling with $K^{3/2}$ at all magnetic fields, which implies that 2D FM fluctuations are dominant above $\sim 20$~K across the whole phase diagram. This is in good agreement with INS results at low fields \cite{CapognaINS}. It is interesting to note that there is a marked disagreement between $1/T_1T$ and $K$ below $\sim 15$~K and above 3.7~T. Naively, this suggests that the FM fluctuations in this temperature range and field range are quenched and the dynamics are dominated by incommensurate antiferromagnetic (AFM) fluctuations. We shall return to the issue of AFM fluctuations later.

 %%%%%%%%%%%%%%%%%%%%%%%%%% Fig2 %%%%%%%%%%%%%%%%%%%%%%%%%%%%%%%%%
\begin{figure}[t]
  \begin{center}
  \includegraphics[width=\linewidth,clip]{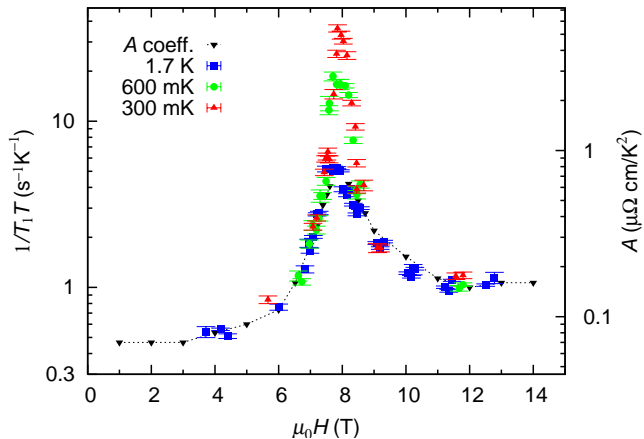}
  \end{center}
  \vskip -0.05\linewidth
\caption{\label{fig:t1vsh1} NMR relaxation rate $1/T_1T$ at various temperatures and the $T^2$ coefficient of the resistivity, $A$ \cite{GrigeraQCP}, versus field.
 }
\end{figure}
%%%%%%%%%%%%%%%%%%%%%%%%%%%%%%%%%%%%%%%%%%%%%%%%%%%%%%%%%%%%%%%%%
%%%%%%%%%%%%%%%%%%%%%%%%%%%%% Fig3 %%%%%%%%%%%%%%%%%%%%%%%%%%%%%%%%%%%%%%%%
\begin{figure}[htbp]
 \begin{center}
 \includegraphics[width=\linewidth,clip]{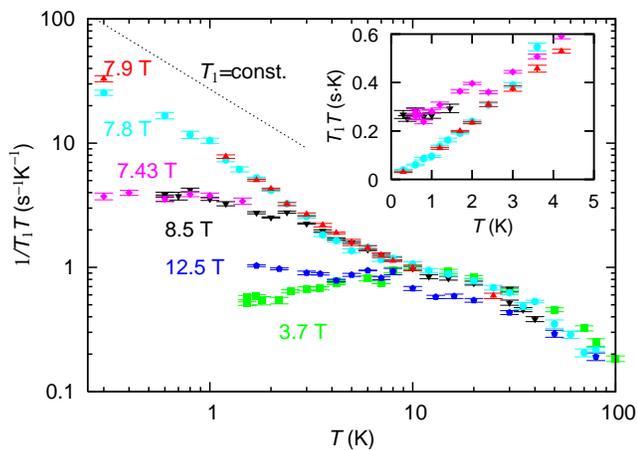}
 \end{center}
  \vskip -0.05\linewidth
\caption{\label{fig:t1vst} $1/T_1T$ vs. $T$ at various magnetic fields. The inset shows a linear-linear plot for the inverse of $1/T_1T$ for $H \sim H_\text{M}$.
}
\end{figure}
%%%%%%%%%%%%%%%%%%%%%%%%%%%%%%%%%%%%%%%%%%%%%%%%%%%%%%%%%%%%%%%%%%%%%%%%%%%%
%%%%%%%%%%%%%%%%%%%%%%%%%%% Fig4 %%%%%%%%%%%%%%%%%%%%%%%%%%%%%%%%%%%%%
\begin{figure}[b]
  \begin{center}
  \includegraphics[width=\linewidth,clip]{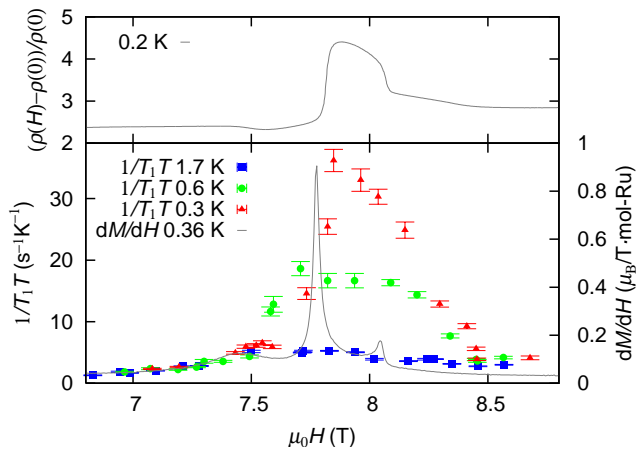}
  \end{center}
  \vskip -0.05\linewidth
\caption{\label{fig:t1vsh2} Comparison of in-plane magnetoresistance ($T=0.2$~K), $1/T_1T$ and uniform susceptibility $dM/dH$ ($T=0.36$~K) \cite{PerryMforcaxis} as a function of magnetic field.
}
\end{figure}

Next, we discuss nature of the metallic state at low temperatures.  In a conventional Fermi liquid, the Korringa relation $1/T_1T \propto N_\text{eff}(E_\text{F})^2$ should be valid.  Here, $N_\text{eff}(E_\text{F})$ is the density of states at the Fermi level and is a measure of the quasiparticle mass enhancements. Previous transport measurements \cite{GrigeraQCP} demonstrated a divergence in the quasiparticle mass as $H \rightarrow H_\text{M}$ by measuring $A$, the $T^2$ coefficient of the resistivity. The $A$ parameter is the strength of the electronic $T^2$ scattering rate and is related to the square of the quasiparticle mass. Thus we expect a new relationship that $1/(T_1TA)$ is independent of $T$ and $H$ in the metallic state at low temperatures. Figure~\ref{fig:t1vsh1} shows $1/T_1T$ and $A$ versus $H$ at 1.7, 0.6 and 0.3~K and up to 14~T. The $A$ data were taken from Ref.~\onlinecite{GrigeraQCP}. The first point to note is that there is good scaling between $A$ and $1/T_1T$ at fields away from $H_\text{M}$. This supports the assertion that the ground state is a Fermi-liquid in this field range, and is in good agreement with the observation of dHvA oscillations and a $T^2$ scattering rate above and below $H_\text{M}$ \cite{PerrySdH,BorzidHvA}. We also observe a divergence in $1/T_1T$ as $H_\text{M}$ is approached, indicating that critical fluctuations are indeed present in the system. These critical fluctuations are renormalized into the effective mass of the quasiparticles in this field range. The temperature dependence of $1/T_1T$ at various fields is shown in Fig.~\ref{fig:t1vst}. At 3.7~T, the broad peak at 15~K is clearly observed and at low temperatures $1/T_1T$ is temperature independent, as previously noted. As $H \rightarrow H_\text{M}$, the temperatures below which $1/T_1T$ is const. decreases, indicating the suppression of the Fermi-liquid state as a function of magnetic field.  Again, this in good agreement with previous experimental data \cite{GrigeraQCP,BorzidHvA}.  Our data represent good microscopic evidence that this systems displays a field tuned critical end point.   
The high field and low field Fermi liquid states also appear to be similar: the renormalized effective masses for $H \gg H_\text{M}$ is only 1.5 times as much as that for $H \ll H_\text{M}$.  This in accordance with the assertion that this is a non-symmetry breaking QCP and is in sharp contrast with a QCP induced by a second-order transition.  One notable example of a second order QCP is the heavy-fermion compound YbRh$_2$Si$_2$ which undergoes an AFM transition at 70~mK \cite{YbRh2Si2NFL,YbRh2Si2muSR}. The application of magnetic fields suppresses the antiferromagnetic transition temperature to zero Kelvin at low fields.  The $A$ coefficient shows striking asymmetric behavior with respect to the QCP (the difference is more than a factor of 5) \cite{YbRh2Si2Acoeff}, suggesting that the character of the Fermi-liquid states is quite different \cite{QCPnHall}.  We continue with a discussion on the nature of the critical fluctuation close to the critical field. 

It can be seen from Figs.~\ref{fig:t1vsh1} and \ref{fig:t1vst}, that the Korringa relation dramatically collapses close to $H_\text{M}$; $1/T_1T$ is not temperature independent down to 0.3~K and hence the system cannot be described as a Fermi-liquid.  Strikingly, $1/T_1T$ at $H_\text{M} \sim 7.9$~T continues to increase down to 0.3~K. When $T_1T$, related to the inverse of $\sum_{\bm q} \chi''(\bm q, \omega_\text{res})$ is plotted against $T$ as shown in the inset of Fig.~\ref{fig:t1vst}, it is interesting to note that spin fluctuations at 7.9~T have a singularity at $T \sim 0$.  This suggests that the spin fluctuations possess a \textit{quantum} critical character. To check quantitatively how close to the \textit{quantum} critical point this system resides, we have fitted $1/T_1T$ to a modified Curie-Weiss law, 
% \begin{equation}
$
 1/T_1 T \propto \left(T^\alpha - \mathit{\Theta}^\alpha\right)^{-1}
$ \cite{CeCu6AuINSNature}.
% \end{equation}
At 7.9~T, we obtain $\mathit{\Theta} = 0.01 \pm 0.04$~K and $\alpha = 1.1 \pm 0.04$ from the fitting below 5~K. The very small $\mathit{\Theta}$ indicates that the criticality in this material is purely \textit{quantum} as opposed to thermal in nature to within the experimental errors.  In addition, $\alpha \sim 1$  implies $T_1 =$const. at least over 1.5 decades of temperature (0.3~K~$< T <$~10~K) indicating that the quantum critical fluctuations dominate the properties of the system below 10~K at $H_\text{M}$.
The temperature independent behavior of $1/T_1$ also suggests that the dynamical susceptibility $\chi''(\bm q_i,\omega)$ at the dominant wave vector $\bm q_i$ is scaled to $\sim \omega/T$ in the small $\omega$ region. This represents the first microscopic evidence of MMQC in this system. 

As we mentioned earlier, a MMQCP is associated with a field-induced quantum critical Stoner instability, which suggests that the fluctuations should exhibit a FM character.  We have investigated this hypothesis by comparing $1/T_1T$ with uniform susceptibility $dM/dH$ \cite{PerryMforcaxis} as shown in Fig.~\ref{fig:t1vsh2}.  Away from $H_\text{M}$ in the Fermi-liquid region, $dM/dH$ and $1/T_1T$ are in reasonable agreement. However, close to $H_\text{M}$ there is a marked difference.  Multiplepeaks seen in $dM/dH$ were not observed in $1/T_1T$ with in the experimental accuracy. Furthermore, $1/T_1T$ exhibits a maximum not at metamagnetic steps in $M$ (the peaks in $dM/dH$) but in the dome-like region of enhanced scattering in the magnetoregistance \cite{PerrySdH}. This enhancement of $1/T_1T$ suggests that the quantum critical fluctuations are not FM but AFM in nature at the critical field. Indeed, the recent INS experiments at $\mu_0 H = 7.95$~T ($T > 2$~K) revealed that the AFM fluctuations at a small $\bm q$ wavevector $(0,0.2,0)$ have a lower-energy excitations than the FM ones \cite{INSunpub}, which is consistent with our NMR results. Therefore, we consider that the critical behavior observed in $1/T_1T$ at MMQCP is intimately related with these incommensurate AFM fluctuations. It is interesting that the AFM fluctuations, which are the extra degree of freedom other than the order parameter of the metamagnetism, are enhanced and display the quantum critical behavior.  One plausible answer is that since itinerant electron metamagnetism is a field-induced instability of the Fermi surfaces driven by FM fluctuations, the nested AFM fluctuations could couple naturally to the quantum critical fluctuations of the Fermi surfaces.  Thus the non-Fermi-liquid behavior observed in various quantities would be attributable to the intense scattering of the quasiparticles due to the strong Fermi surface instability. 

Finally, we discuss the evidence from our data for the putative phase formation close to $H_\text{M}$.  There appears to be contrasting behavior between $1/T_1T$ and the bulk properties: $1/T_1T$ continues to increase down to the lowest temperatures, whereas the resistivity and the $M(T)$ show a kink at 1~K \cite{PerrySdH, GrigeraDisorder}. From the bulk anomalies, it was recently proposed that Pomeranchuk phase is stabilized around $H_\text{M}$ \cite{GrigeraDisorder}. We point out that the absence an anomaly in $1/T_1T$ at 1~K can be understood if we assume that NMR only detects the microscopic magnetic properties inside the putative magnetic domains and/or that the spectral weight of the relevant fluctuations is so small that the anomaly of $1/T_1T$ is masked by the experimental precision. At the moment, the origin of the difference is not identified. The $T$-dependence of AFM-fluctuation intensity by INS experiments may give an important clue to clarify this issue. In whichever case, the continuous increase in $1/T_1T$ indicates that quantum critical fluctuations persist to the lowest temperatures. 

%\section{Conclusion}
In summary, we have been successful in observing  $^{17}$O-NMR spectra in \bilayer. At high temperatures we have microscopic evidence for the dominance of 2D FM fluctuations. At low temperatures, a Fermi-liquid state is observed except in the vicinity of the MMQCP. Near the MMQCP, $1/T_1$ is $T$-independent down to 0.3~K, indicating that the spin fluctuations possess a quantum critical character. From comparison with the static susceptibility we conclude that these quantum critical fluctuations are most likely AFM and not FM in nature, although the FM fluctuations are thought to be the driving mechanism behind the metamagnetism. 

\begin{acknowledgments}
 We thank S.~Nakatsuji, H.~Yaguchi, H.~Ikeda, K.~Yamada, E.~M.~Forgan, and S.~Hayden for valuable discussions. We particularly acknowledge A.~J.~Schofield, S.~A.~Grigera and A.~P.~Mackenzie for valuable discussions and reading the manuscript before submission. This work was in part supported by the Grants-in-Aid for Scientific Research from JSPS and MEXT of Japan, and by the 21COE program ``Center for Diversity and Universality in
Physics'' from MEXT of Japan.
\end{acknowledgments}

%\newpage Just because of unusual number of tables stacked at end

\bibliography{resubmit.bib} % Produces the bibliography via BibTeX.
\end{document}